\documentclass[12pt]{iopart}
\usepackage{graphicx}
\begin{document}

\title{Spontaneous Formation of Dynamical Groups in an Adaptive Networked System}

\author{Menghui Li$^{1,2}$, Shuguang Guan$^{3,1,2}$, C.-H. Lai$^{2,4}$}

\address{$^1$ Temasek Laboratories, National University of
Singapore, 117508, Singapore}
\address{$^2$ Beijing-Hong Kong-Singapore Joint Centre for Nonlinear
and Complex Systems (Singapore), National University of Singapore,
Kent Ridge, 119260, Singapore}
\address{$^3$ Institute of Theoretical Physics and Department of
Physics, East China Normal University, Shanghai, 200062, People's
Republic of China}
\address{$^4$
Department of Physics, National University of Singapore, Singapore
117542}

\begin{abstract}
In this work, we investigate a model of an adaptive networked dynamical
system, where the coupling strengths among phase oscillators
coevolve with the phase states. It is shown that in this model the
oscillators can spontaneously differentiate into two dynamical
groups after a long time evolution.  Within each group, the oscillators
have similar phases, while oscillators in different groups have
approximately opposite phases. The network gradually
converts from the initial random structure with a uniform distribution
of connection strengths into a modular structure which is characterized
by strong intra connections and weak inter connections. Furthermore,
the connection strengths follow a power law distribution, which is a
natural consequence of the coevolution of the network and the
dynamics. Interestingly, it is found that if the inter connections
are weaker than a certain threshold, the two dynamical groups will
almost decouple and evolve independently. These results are helpful
in further understanding the empirical observations in many
social and biological networks.
\end{abstract}

\maketitle


Modularity frequently occurs in many social and biological networked
systems \cite{Nodular:empirical}, which is generally believed to
correspond to certain functional groups \cite{functional}. Usually,
in modular networks, the intra connections are stronger than the
inter connections
\cite{CN:Mobile,Evo:Community,May1973,foodweb:weak}. However, they
both play important roles in maintaining the network structure and
functions. P. Csermely pointed out that the strong links can define
the system, while the weak links are crucial to the stabilization of
complex system \cite{Link:weak}. Such examples can be found in many
situations, such as the connectivity of social networks
\cite{CN:Mobile}, group survival \cite{survival}, social efficiency
\cite{socialefficiency}, firm efficiency \cite{firmefficiency}, and
ecosystem stability \cite{ecosystem}. Furthermore, in many networks,
such as the natural food webs \cite{foodweb},  mobile networks
\cite{CN:Mobile}, author collaboration networks \cite{coauthor},
metabolic netwoks\cite{E.coli} and neural networks\cite{neural}, it
is found that most of interactions are weak, and only a few
interactions are strong, which usually leads to a power law
distribution of the connection strengths \cite{CN:Mobile,coauthor,
E.coli,neural}.

In the past decade, there are extensive works exploring networked
complex systems. Mainly, these works focus on either the topological
structures of the networks \cite{CN:REV}, or the dynamics on the
networks \cite{SYN:REV}. Nevertheless, in various realistic systems,
especially the biological and social systems, in principle the
network topology and dynamics are strongly dependent on each other.
Thus any network structures and dynamical patterns that emerged are
actually the results of the coevolution of the network dynamics and
topology \cite{adaptive:REV}. For example, the change of the
synaptic coupling strength between neurons depends on the relative
timing of the presynaptic and postsynaptic spikes in neural networks
\cite{STDP}, and in the mobile communication networks
\cite{CN:Mobile}, the connection strengths are determined by the
dynamical behaviors of the mobile agents.

Recently attentions have  been paid to the adaptive coevolutionary
networks \cite{adaptive:REV}.  These include the  adaptive rewiring
links \cite{adaptive:rewir,adaptive:community},  and the adaptive
altering connection strength
\cite{adaptive:group,adaptive:SYN,adaptive:weight,adaptive:neural}
based on the states of local dynamics.
However, the previous studies still focus mainly on the topological
properties of the networks, while neglecting the dynamical evolution and
characteristics, which are actually one very important aspect of
networked  dynamical systems. We noticed that in many social and
biological networked systems, with the evolution of the network
topology, dynamically the system may form different functional
groups corresponding to different dynamical states. One such example
is the mammalian brain, in which the connections are plastic
\cite{STDP}. It is known that the mammalian brain is composed of a
number of functional groups, within which the nodes can be regarded
as sharing similar dynamical states. However, so far, how the
dynamical groups are generated during the coevolution of network
structure and dynamics has not been investigated from the point of
view of complex networks.

Motivated by this idea, in the present work, we set up a toy
adaptive network model consisting of phase oscillators. Due to the
simplicity of the dynamics,  phase oscillators have been frequently
used to describe many simplified real dynamical systems, such as
biological networks, chemical oscillators and so on
\cite{REV:Kurmoto}. In our model, the coupling functions adopt the
higher order Fourier modes, and the connection strengths are coupled
with the local dynamical states following the plasticity function.
Particularly, we investigate what kinds of the dynamical states and
network structures can be formed as the result of the coevolution of
both network dynamics and topology. Mainly, our study presents three
new results. (i) The dynamical groups can be spontaneously formed in
our model, i.e., in-phase and anti-phase synchronized states
simultaneously exist in our system. In the previous work
\cite{neuronmodel}, though the desynchronized states and the
synchronized states coexist and are both stable, the network only
tends to be one of the two states, depending on its initial mean
coupling. While in our model, the oscillators within (between)
groups tend to in-phase (anti-phase) synchronization. (ii) The
connection strengths in the network can self-organize into a power
law distribution from the initial random distribution. In addition,
communities, which correspond to the dynamical groups in our model,
can also be spontaneously formed. The community structure and the
power law distribution of the connection strengths are common in
many empirical networked systems. (iii) The resource constraint can
significantly affect the formation of the dynamical groups. If the
total connection strength is a finite constant, the network tends to
split into two dynamical groups: within each group the oscillators
are in-phase synchronized, while the oscillators in different groups
are anti-phase synchronized. However, if there is no resource
constraint, the two groups finally merge into one.

In our model, the dynamical equations for the networked phase
oscillators read:
\begin{equation}
\dot{\theta}_m = \omega_m + \gamma \sum_{k=1}^{N}w_{mk}
\Gamma(\theta_k - \theta_m). \label{model}
\end{equation}
Here, $m,k=1,2,\ldots,N$ are the oscillator (node) indices, and
$\gamma$ is the uniform coupling strength. $\theta_m$ and $\omega_m$
are the instantaneous phase and intrinsic frequency of the $m$th
oscillator, respectively.
 $W=\{ w_{mk} \}$($w_{mk}=w_{km}$) is the
weighted coupling matrix, where $w_{mk}>0$ is the coupling strength
if nodes $m$ and $k$ are directly connected, and $w_{mk}=0$
otherwise. In order to  generate possible dynamical groups in our
model, we tentatively choose the coupling function $\Gamma(\phi)$ as
the higher order of Fourier modes, i.e.
 $\Gamma(\phi)=\sin(h\phi)$ ($h=2,3,4,\cdots$) \cite{muticluster},
where the parameter $h$ can control the number of groups. Without
losing generality, we set $h=2$ throughout this paper.

In the coevolutionary networked system, how the network topology
couples with the dynamics is crucial to both the dynamical pattern
and topological structure that result.  In our model, we propose a
coupling rule  for the connection strength $w_{mk}$ based on
following hypothesis: $w_{mk}$ is a finite real number, and the
connections will be strengthened (weakened) if the phase differences
are smaller (bigger) than some threshold $\alpha$. Actually, this
can be regarded as an extension of the spike-timing dependent
plasticity (STDP) rule \cite{STDP}. In fact, in many realistic
networked systems, individuals with similar states usually tend to
form the group
 which has relatively stronger intra connections inside.
 For instance, in human society,
individuals with similar attributes are easily organized into the
same communities
\cite{Evo:Community,empirical:email,review:socialnetwork}.
Meanwhile, similarity will breed connection
\cite{review:socialnetwork}, indicating that the relations among
individuals with similar attributes may be constantly strengthened,
while those among individuals with dissimilar attributes may be
gradually weakened. Based on the above consideration, the change of
the connection strength is assumed to satisfy the following
equation:
\begin{equation}
\frac{dw_{mk}}{dt} = \epsilon w_{mk}\Theta
(\Delta\theta_{mk},\alpha) \Lambda(\Delta\theta_{mk}),
\label{linkmodel}
\end{equation}
where $\Delta\theta_{mk}=|\theta_k-\theta_m|$ ($0\leq
\Delta\theta_{mk} \leq \pi$) is the phase difference between
oscillators $m$ and $k$. $w_{mk}$ in the right hand side of the
equation ensures that the rate of change rate of the link weight is proportional
to itself, and $w_{mk}\geq 0$ always. $\epsilon$ is a constant which
can be chosen to make the time scale
 of the network topology evolution
much longer than that of the local dynamics of the oscillators. The
function $\Theta(\phi, \alpha)$ determines how the coupling strength
evolves according to the phase difference between oscillators. In this
study, we set $\Theta(\phi, \alpha)=e^{-2|\phi-\pi/2|}$. The
function $\Lambda(\phi)$, which is similar to the sign function,
controls either the strengthening or weakening of the connections based on
the phase differences. For simplicity, we assume the form
$\Lambda(\phi)=\Gamma(\phi)$. The form $\Theta (\phi, \alpha)
\Lambda(\phi)$ is similar to the STDP rule, which has been widely
used in neural network studies \cite{STDP,neuronmodel}. The
difference is that the STDP rule\cite{STDP,neuronmodel} depends on the
relative timing $\Delta t$ of presynaptic and postsynaptic spikes
and the critical window $\tau$, while the plasticity function in our
model depends on the phase difference $\Delta \theta$ and the
connection strength itself. In addition, the exponential function
$\Theta(\phi, \alpha)=e^{-2|\phi-\pi/2|}$ is modulated by the sine
function $\Lambda(\phi)=\sin(2\phi)$, which makes the plasticity
function not a monotone function on the same side of the threshold
value, e.g., $\Delta \theta <\pi/2$.

With the above assumptions, the model is fully described by
\begin{eqnarray}
\begin{array}{l}
\dot{\theta}_m = \omega_m + \gamma \sum_{k=1}^{N}w_{mk}
\sin[2(\theta_k - \theta_m)] \\
\\
\frac{dw_{mk}}{dt} = \epsilon
w_{mk}e^{-2|\Delta\theta_{mk}-\pi/2|}\sin(2\Delta\theta_{mk}).
\end{array}\label{models}
\end{eqnarray}
In this study, the natural frequencies and initial conditions of the
oscillators are chosen randomly from the range $[0,1]$ and $[-\pi,
\pi]$, respectively. It is known that in many practical adaptive
networks, the ``resource", which can be represented by the
summation of all connection strength in the network, is usually limited.
Consequently, all connections will compete for this resource.
Therefore, in our model we define the ``resource" as $M=L \langle
w\rangle$, where $L$ is the number of total connections and $\langle
w\rangle$ is the average connection strength. In our simulation, we
use the normalization$\langle w\rangle=1$ during the evolution in order to make
the  ``resource" $M=L$, i.e.,  the total ``resource" to be allocated
is a constant during evolution.

The collective behavior of the dynamical system can be conveniently
described by two order parameters $R$ and $F$. The order parameter
$R$, which characterizes whether the global coherence occurs or not,
is defined as
\begin{equation}
R= \frac{|\sum_{m=1}^N s_me^{i\theta_m}|}{\sum_{q=1}^N
s_q},\label{globalorder}
\end{equation}
where $s_m$ is the strength of node $m$, i.e. $s_m=\sum_k w_{mk}$.
This type of order parameter has been widely used to characterize
the phase synchronization in complex network \cite{rglobal}. From
the definition in papers \cite{rglobal}, it seems natural to use Eq.
(\ref{globalorder}) as the order parameter in weighted networks. The
order parameter $F$, which measures the fraction of all link weights
synchronized in networks \cite{rlink}, is defined as
\begin{equation}
F=
\frac{|\sum_{mk}w_{mk}e^{i(\theta_m-\theta_k)}|}{\sum_{lq}w_{lq}}.\label{localorder}
\end{equation}
In adaptive oscillator networks where the connections are coupled
with the dynamical states, the order parameters $R$ and $F$ can be
jointly used to characterize whether the local coherence within
subnetwork takes place. For example, if $R\approx 0$ and $F\gg R$
after a long time evolution from random initial phases on random
networks, it indicates that the local synchronization (rather than
the global synchronization) emerges within subnetworks, i.e., the
dynamical groups have been generated in the system.


First, we consider a simplified situation: a two-oscillator system.
In this case, the dynamics can be rewritten in terms of two
variables, $\Delta \theta=(\theta_1-\theta_2)$ and $w$, as
\begin{eqnarray}
\begin{array}{l}
\frac{d\Delta\theta}{dt}= \omega_1-\omega_2 - 2\gamma w\sin(2\Delta \theta) \\
\\
\frac{dw}{dt} = \epsilon
we^{-2|\Delta\theta-\pi/2|}\sin(2\Delta\theta).
\end{array}\label{twoosc}
\end{eqnarray}
From the above equations, we can see that if $|\Delta
\omega|=|\omega_1-\omega_2|=0$, the system will have stable
equilibrium states $\Delta\theta^* =0$ or $\pi$, and the final
connection strength $w^*$ is a finite constant. These two states
correspond to the in-phase synchronization and the anti-phase
synchronization of the two oscillators, respectively. If $|\Delta
\omega| \neq 0$, strictly speaking the two-oscillator system does not have
any equilibrium states. This implies that the coupling strength will
always be varying during the evolution. Nevertheless, if the rate of change
of the connection strength is much slower than the phase
dynamics, we can approximately regard $w$ as a constant. In this
case, we can obtain the stable equilibrium states of $\Delta\theta$
provided that $|\Delta \omega|\leq 2\gamma w$, i.e.,
\begin{equation}
\Delta\theta^{*}=\left\{
\begin{array}{l}
 \frac{1}{2}\arcsin|\frac{\Delta\omega}{2\gamma w}|\\

 \pi-\frac{1}{2}\arcsin|\frac{\Delta\omega}{2\gamma w}|.
\end{array}\right.
\end{equation}
In our numerical simulations, the above analysis has been verified.

\begin{figure}[tbp]
\begin{center}
\includegraphics[width=0.8\linewidth]{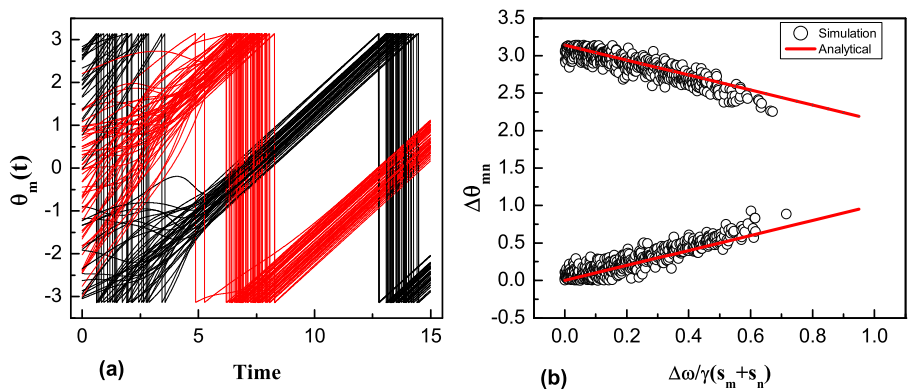}
\caption{(Color online)(a) The time evolution of the oscillator
phases. After transients, oscillators spontaneously differentiate
into two dynamical groups with different states. (b) The comparison
between the analysis and the simulation results of phase differences
among oscillators. The network parameters are $N=100$, $\langle
k\rangle=20$, $\gamma=0.04$, and $\epsilon=0.01$. Initially
$\omega_m\in [0,1]$ and $w_{mn}=1$.}\label{fig1}
\end{center}
\end{figure}

Next, we consider the case of a many-oscillator system. Without losing
generality, the initial network topology is chosen as a random
structure, and the initial connection strengths are chosen uniformly
from the range $(0,2]$. To monitor the evolution, we record the
instantaneous phases of all the oscillators ($\{\theta_m(t)\}$).
Interestingly, it is found that after the transient period, the
oscillators can spontaneously separate into two dynamical groups.
Within each group, all oscillators have similar phases. Meanwhile, the
two dynamical groups as a whole tend to approximate anti-phase
synchronization as shown in Fig. \ref{fig1}(a). Through extensive
numerical simulations, we found that the sizes of the two groups
depend on the initial conditions. In general, they are almost equal
to each other when the initial phases are chosen uniformly. Of
course, if all the oscillators are identical, the coevolution can still
generate two dynamical groups as nonidentical system. In this case,
the phase states within each group are strictly identical.

The collective behavior of the dynamical system with multiple
dynamical groups can also be described by the following parameter
\cite{adaptive:weight},
\begin{equation}
R'= \frac{|\sum_{m=1}^N s_me^{i2\theta_m}|}{\sum_{q=1}^N
s_q}.\label{multipleorder}
\end{equation}
If the order parameter $R'$  converges to $1$ and the order
parameter $R$ converges to $0$, this also implies that dynamical
groups have formed. The difference between $F$ and $R'$ is that $F$
can characterize the properties of the dynamical states and the
topology of weighted networks simultaneously, while $R'$ can mainly
characterize the properties of the dynamical states. In order to
explain the formation of different dynamical groups in our model Eq.
(\ref{model}), we rewrite it in a more convenient form by defining
the local order parameter according to Eq. (\ref{multipleorder})
\begin{equation}
r'_{m}e^{i2\psi_m}=\frac{1}{s_m}\sum_{k=1}^{N}w_{mk} e^{i2\theta_k}.
\label{localparameter}
\end{equation}
Here $r'_m$ with $0<r'_m<1$ measures the local coherence among the
neighbors of oscillator $m$.  $\psi_m$ is the average phase, and
$s_m$ is the strength of node $m$, i.e. $s_m=\sum_k w_{mk}$. With
this definition, Eq. (\ref{model}) becomes
\begin{equation}
\dot{\theta}_m = \omega_m + \gamma r'_m s_m
\sin[2(\psi_m-\theta_m)]. \label{mmodel}
\end{equation}
When $\gamma\rightarrow 0$, Eq.(\ref{mmodel}) yields
$\theta_m\approx \omega t+\theta_m(0)$, that is, the oscillators
evolve according to their own natural frequencies. The oscillators
are neither in-phase nor anti-phase synchronized, i.e.
$r'_m\rightarrow 0$ as $t \rightarrow \infty$. On the other hand, in
the limit of strong coupling, the oscillators tend to anti-phase
synchronized, $r'_m\rightarrow 1$ and $2\psi_m -2\theta_m \approx
2q_m\pi$($q_m=0,\pm 1$), i.e. $2\psi_m -2\theta_m - 2q_m\pi\approx
0$. Consequently, Eq. (\ref{mmodel}) can be rewritten as
\begin{equation}
\dot{\theta}_m = \omega_m + 2\gamma s_m (\psi_m'-\theta_m),
\label{linearmodel}
\end{equation}
where $\psi_m'=\psi_m-q_m\pi$. Thus, the phase difference
$\Delta\theta_{mn}=\theta_m- \theta_n$ between $m$ and $n$ becomes
\begin{equation}
\frac{d\Delta\theta_{mn}}{dt} = \omega_m-\omega_n + 2\gamma[ s_m
(\psi_m'-\theta_m)-s_n (\psi_n'-\theta_n)]. \label{phasedifference}
\end{equation}
From $\frac{d\Delta\theta_{mn}}{dt}=0$, we can obtain the
equilibrium value $\Delta\theta_{mn}$, i.e.,
\begin{equation}
\Delta\theta_{mn}=\frac{\Delta\omega_{mn}}{\gamma(s_m+s_n)}+\psi_m'-\psi_n'+\frac{s_m-s_n}{s_n+s_m}(\psi_m'-\theta_m+\psi_n'-\theta_n),
\label{difference}
\end{equation}
where $\Delta\omega_{mn}=\omega_{m}-\omega_{n}$, and
$\frac{s_m-s_n}{s_n+s_m}(\psi_m'-\theta_m+\psi_n'-\theta_n)$ is the
high-order infinitesimal, which could be neglected. When oscillators
$m$ and $n$ tend to in-phase (anti-phase) synchronization,
$\psi_m'-\psi_n'\approx q\pi$ ($q=0, \pm 1$), so the equilibrium
values of the phase difference $\Delta\theta^{*}_{mn}$ are
\begin{equation}
\Delta\theta^{*}_{mn}=\left\{
\begin{array}{l}
 \frac{|\Delta\omega_{mn}|}{\gamma(s_m+s_n)}\\

 \pi-\frac{|\Delta\omega_{mn}|}{\gamma(s_m+s_n)}.
\end{array}\right.
\end{equation}
As shown in Fig. \ref{fig1} (b), our numerical simulations of the
phase differences are consistent with the analytical
results.


Physically, the spontaneous formation of two different dynamical
groups in  our model can be attributed to the adaptive evolution
rule described by Eq. (\ref{linkmodel}). Based on this equation, the
connection strength among oscillators with initially close phases
will be enhanced. Meanwhile, if two oscillators initially have large
phase difference (e.g., $\Delta \theta>\pi/2$), the connection
strength between them will be weakened during evolution. As a
combined effect of these two ``forces", the networked oscillators
self-organize into two dynamical groups after a long time evolution.
Within the same group, the oscillators have similar states, while
oscillators in different groups have approximate anti-phases.
Interestingly, in many social and biological systems, we often find two
groups are formed with opposite states. For instance, in human society,
individuals with homogeneous character, e.g., the same generation or
living in the same neighborhood, are disposed to associate
\cite{Evo:Community}, and conflicting (accordant) characters could
weaken (strengthen) the social contacts. In food webs, if the living
habits of predator and prey are similar (different), the
predator-prey relationships are strong (weak) \cite{foodweb:weak}.
Our model thus can shed light on the origin of the formation of such
dynamical groups.

\begin{figure}[tbp]
\begin{center}
\includegraphics[width=0.8\linewidth]{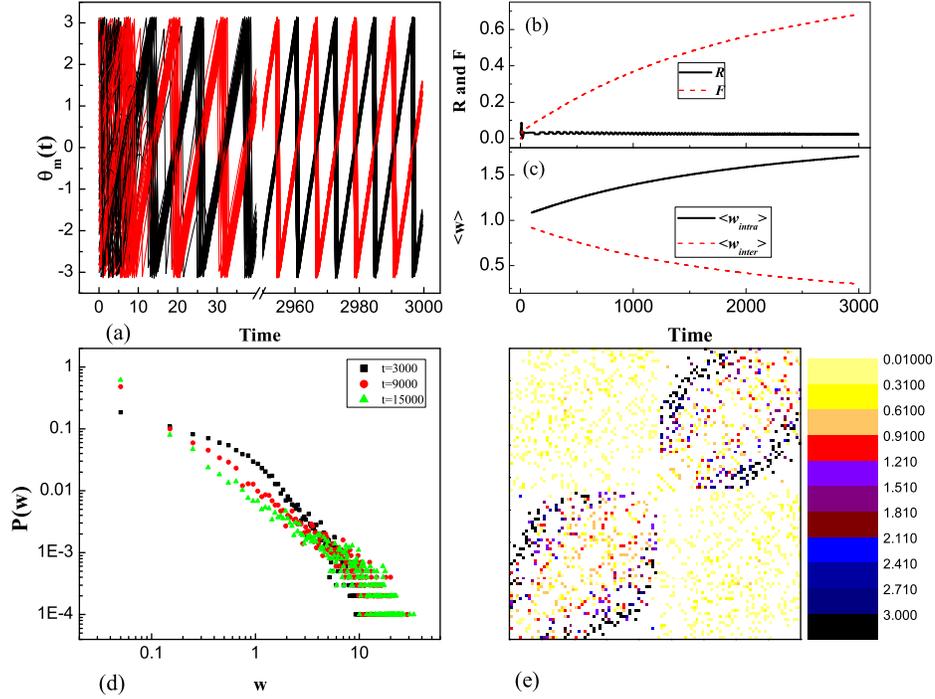}
\caption{(Color online) Characterization of the formation of the
dynamical groups and the modular structure of the network. (a) The
evolution of the oscillator states. (b) The evolution of the order
parameters, where $F$ keeps increasing, but $R$ always maintains
very small values, indicating that the dynamical groups have formed.
(c) The evolution of the average connection strength, where the
average strength of the inter connections ($\langle
w_{inter}\rangle$) decreases all the time, while the intra
connection strength ($\langle w_{intra}\rangle$) keeps increasing.
(d) The distribution of the connection strength for the network at
t=3000, 9000 and 15000. The longer is the time t, the more obvious
is the power law distribution of connection strength. This result is
the average of $20$ runs with different initial conditions. (e) The
snapshot of weight matrix $w_{mk}$ at t=3000, where modular
structure occurs simultaneously with the formation of dynamical
groups. The indices of the oscillators have been rearranged
according to the phase. The parameters are same as those of Fig.
\ref{fig1}, except for $w_{mn}\in(0,2]$ initially. } \label{fig2}
\end{center}
\end{figure}

With the formation of dynamical groups, how the network structure
evolves is another important question. In this work, we do not
consider the rewiring of network connections. Instead, we fix the
network topology and focus on how the network connections compete
for the limited ``resource", i.e., the reallocation of the
connection strength. At every time steps, we normalize $\langle
w\rangle=1$, i.e. $w_{mn}^*=\frac{Mw_{mn}}{\sum_{j>i} w_{ij}}$, in
order to make the  ``resource" $M=L$. In Fig. \ref{fig2}, we
illustrate using one typical example the properties of the network
structure. As shown in Fig. \ref{fig2}(a), the oscillators in the
network self-organize into two dynamical groups with different phase
states, i.e., oscillators within the same group have similar but
nonidentical states, while oscillators in different groups have
approximate anti-phases. The formation of the dynamical groups can
be characterized by the two order parameters $R$ and $F$. As shown
in Fig. \ref{fig2} (b), $F$ keeps increasing during the evolution,
but $R$ always maintains very small values. This suggests that local
dynamical patterns (rather than a global one) gradually form in the
system. To characterize the emerging network structure, we show the
average strength of the inter connections $\langle w_{inter}\rangle$
and the intra connections $\langle w_{intra}\rangle$ as a function
of time in Fig. \ref{fig2} (c). It is evident that the average
strength of the inter connections $\langle w_{inter}\rangle$
decreases, while the intra connection strength $\langle
w_{intra}\rangle$ keeps increasing with time. These results indicate
that with the appearance of the dynamical groups, the distribution
of the connection strengths  in the network also changes. From the
initial random distribution, the connection strengths within the
groups are gradually strengthened, while the connection strengths
between the two groups are weakened simultaneously. In this way,
after a long time evolution, the topological structure of the
networked system has the following characteristics, as shown in
Figs. \ref{fig2} (d)-(e). First, the network evolves into a modular
structure with the formation of dynamical groups. Secondly, it is
found that the network consists of many weak connections and a few
strong connections. Thirdly, to be specific, we have verified that
the distribution of the connection strengths follows a power law, as
compared to the initial random distribution. It should be pointed
out that this power-law distribution of the link weights in the
present model is a natural consequence of the coevolution of the
network topology and dynamics. These results are consistent with the
empirical observations of social systems \cite{CN:Mobile,coauthor},
biological systems \cite{foodweb,E.coli} and neural
network\cite{neural,adaptive:neural}. For instance, in neural
networks\cite{neural,adaptive:neural}, the synaptic strengths of
experimental data follow a power law distribution.

\begin{figure}
\begin{center}
\includegraphics[width=0.8\linewidth]{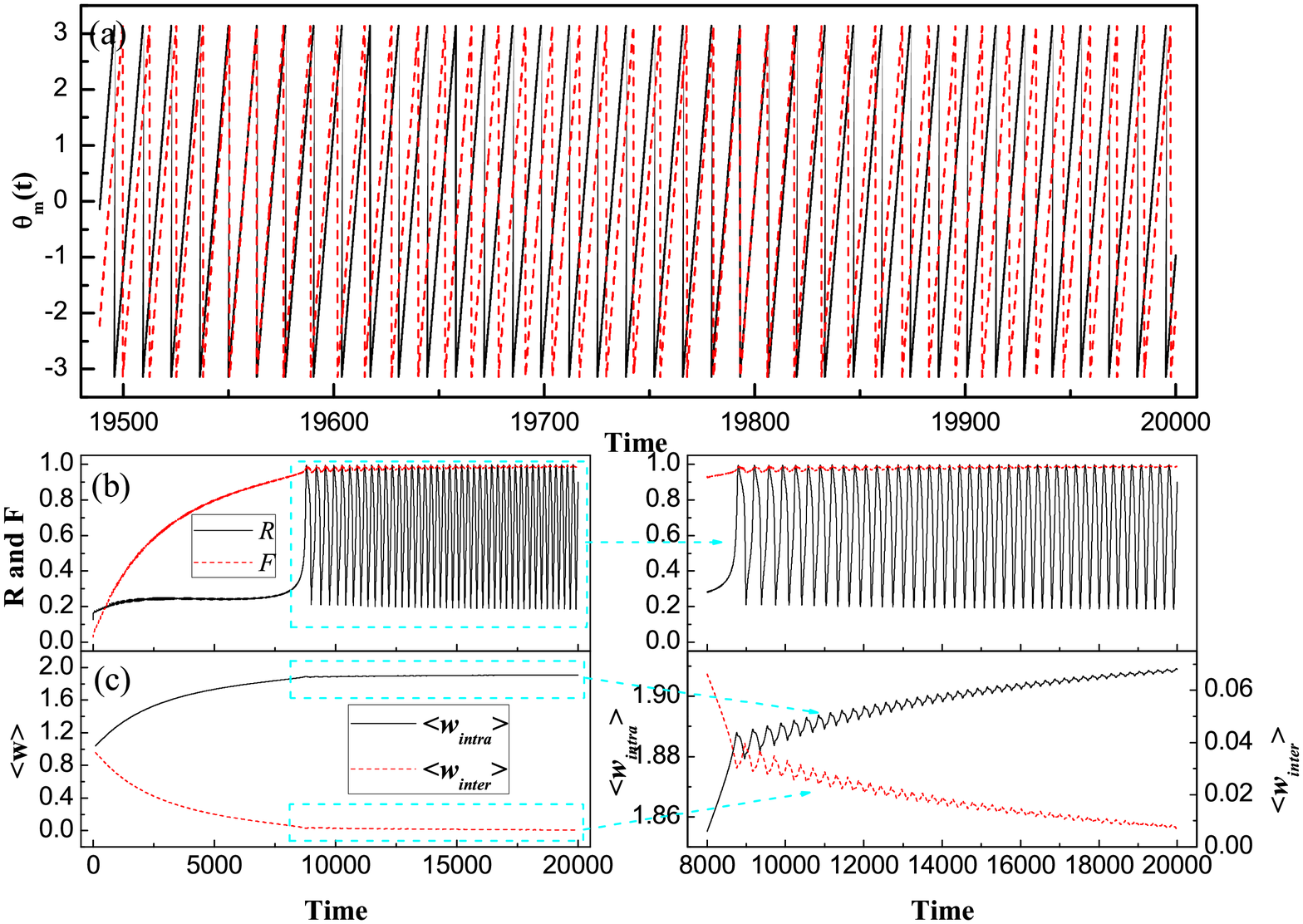}
\caption{(Color online) Characterization of the dynamical and
topological properties of the network after extremely long time
evolution. (a) The schematic diagram of the evolution of the two
groups of oscillators as a whole, where each group of oscillators
behaves like an individual oscillator. (b) and (c) The evolution of
order parameters and the average inter and intra connection
strength, showing when the inter connections become very small, the
two groups of oscillators almost decouple. All the network
parameters are the same as that in Fig. \ref{fig2}.} \label{fig3}
\end{center}
\end{figure}

As shown by Fig. 2, with the evolution of the networked dynamics,
the oscillators begin to separate into two groups with different
states. In Fig. 2(c), it is shown that during the evolution process
the average intra connection strength is gradually enhanced while
the average inter connection strength is weakened always. Here, the
question is: what would the two groups behave when the connections
between them become weak enough? In Fig. 3, we further explore this
situation. Interestingly, it is found that when the inter
connections between the two groups are too weak, e.g., $\langle
w_{inter}\rangle < 0.1$ \cite{explain}, the two dynamical groups
effectively decouple and evolve independently according to their own
frequencies. As shown in Fig. 3(a), the frequencies of the two
groups are almost equal to each other, and during the
evolution, their phases will slowly approach the same value and then
begin to separate. This occurs repeatedly, which lead to
the regular oscillation of the global order parameter $R$ as shown
in Fig. 3(b). Meanwhile, when the phase differences between the two
groups become small enough, according to Eq. (3), the inter
connection strength will be enhanced. However, this trend
will not last for long since the phase differences of the two groups
will begin to be significant soon. As shown in Fig. 3(c), although both
the average inter connection strength and the average intra connection strength each
oscillate with a small range, the trends are an overall decrease for the average inter connection strength and increase for the average intra  connection strength.  This implies that these two dynamical
groups will become more and more independent after a long time
evolution. Moreover, even in the collective oscillatory regime,
the distribution of the link weights still follow a power law (as shown
in Fig. \ref{fig2}(d)).

\begin{figure}[tbp]
\begin{center}
\includegraphics[width=0.8\linewidth]{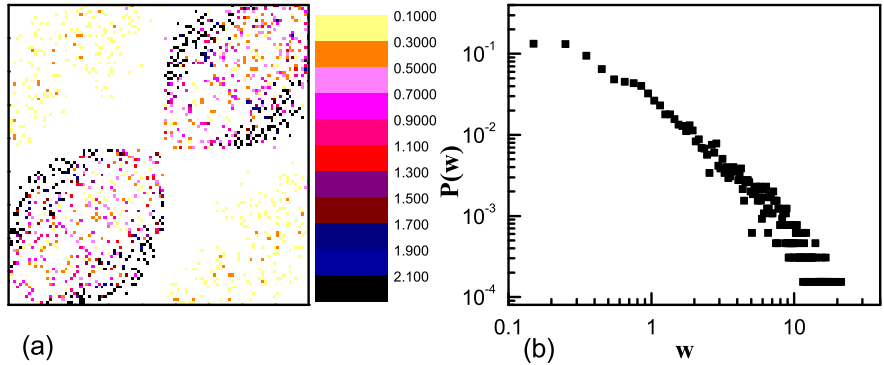}
\caption{(Color online) Characterization of the properties of the
observable networks consisting of the ¡°active¡± connections.  (a)
The weighted matrix of the observable networks, where the modularity
is more distinct when compared with Fig. \ref{fig2}(e). (b) The
distribution of the active connection strengths, which follows a
power law. This result is the average of $20$ runs with different
initial conditions.  All network parameters are the same as that in
Fig. \ref{fig2}} \label{fig4}
\end{center}
\end{figure}

In realistic networked systems, if the connections are extremely
weak, it may be impossible to measure them. As a consequence, any
observed real network should consist of connections whose strengths
are strong enough to be measured. In our model, we found that there
exists a large number of weak links and many of these have no
opportunity to be enhanced again. Therefore, from the practical
point of view they may not be observable at all after a long time
evolution. To distinguish them, we can define the active connections
as follows: if $w_{mk}$ exceeds a threshold value, the connection
between oscillator $m$ and $k$ is regarded as ``active"; otherwise
it is ``inactive". The threshold can be reasonably taken as the
average of the inter connections, i.e., $\langle w_{inter}\rangle$.
Using this criterion, we obtain observable networks after a long
time evolution based on our model. Numerically, we let the networked
system evolve from many different initial conditions. After
$t=5000$, we start taking snapshots of $w_{mk}$. After discarding
the ``inactive" connections, we obtain the observable networks only
consisting of the ``active" connections. It is found that in these
observable networks, the modular property becomes even more
distinct. As shown in Fig. \ref{fig4}, the oscillators can be
reasonably partitioned into two communities, and the distribution of
connection strengths still approximately satisfies the power law
relation. These results suggest that the widely observed community
structure and the power-law distribution of link weights in complex
networks could emerge simultaneously from the coevolution of the
network topology and dynamics.

\begin{figure}[tbp]
\begin{center}
\includegraphics[width=0.5\linewidth]{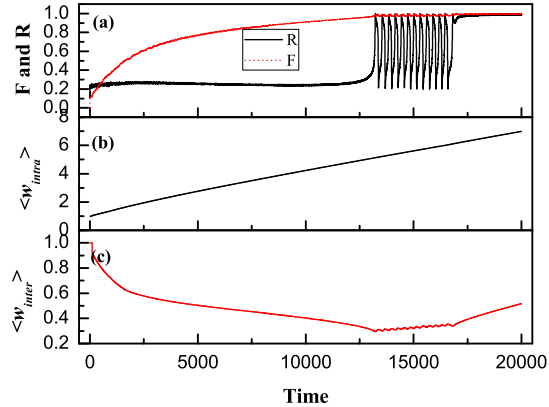}
\caption{(Color online) Characterization of the dynamical and
topological properties of the network after extremely long time
evolution, where the total connection strength is not limited. (a)
The order parameters $F$ and $R$ characterize three distinct stages
of network evolution, i.e., first the dynamical groups have formed;
then the two groups of oscillators almost decouple; and finally all
oscillators achieve in-phase synchronization. (b) The evolution of
the average intra connection strength, which keeps increasing. (c)
The evolution of the average inter connection strength, which first
decreases and then increases. All the network parameters are the
same as that in Fig. \ref{fig2}.} \label{fig5}
\end{center}
\end{figure}

In the above studies, we have limited the total connection strength
as a constant in the network. This consideration makes sense in
certain practical circumstances. For example, the bandwidth of a
local area network in a university is always limited.
However, in other networks, e.g., the
social acquaintance network, there is no need to limit the total connection
strength during the network evolution. In this case, how would the
dynamics and the network structure coevolve? In the following, we
investigate one such example. It is found that the initial stage of
the network evolution is quite similar to the case when the total
network connection strength is limited. As shown in Fig.
\ref{fig5}(a), the global order parameter $R$ remains small while
the local order parameter $F$ keeps increasing. This indicates that
the two dynamical groups have been generated. In Figs. \ref{fig5}(b)
and (c), it is shown that the average intra connection strength continues to
increase, while the average inter connection strength keeps decreasing.
This is just the reason which leads to the formation of the
dynamical groups. When the inter connection strength among the
groups is small enough, the two groups almost decouple and they
behave just like two independent oscillators. However, with the
further increase of time, contrary to the previous situations, the
inter connection strength starts to gradually increase as
shown in Fig. \ref{fig5}(c). Due to this strengthening of the inter
connections, the two dynamical groups eventually will merge into one
and all oscillators will achieve in-phase synchronization.
Therefore, our results suggest that during the evolution of a
network, the limitation of the total connection strength is in favor
of the formation of stable dynamical groups.


In summary, we have investigated a coevolutionary networked model.
In this model,  the node dynamics are described by phase
oscillators, and the connections among oscillators are coupled with
the dynamical states. By adopting a simple evolution rule, it is
shown that the evolution of the networked system naturally leads to
two dynamical groups with different phase states. Simultaneously,
with the formation of the dynamical pattern, the network also
converts from the initial random structure with a uniform
distribution of connection strengths to the final modular network
with a power-law distribution of the connection strengths.
Interestingly, it is found that if the total connection strength is
limited as a constant, the two dynamical groups will almost decouple
eventually when the inter connection is too weak. On the contrary,
if the total connection strength does not have an imposed limit, the
two dynamical groups will finally merge into one with all the
oscillators achieving in-phase synchronization. In our numerical
simulations, the above results have been qualitatively verified on
networks with sizes up to $N=1000$. Although the model studied is
simple, it essentially captures the interplay between the network
topology and dynamics. Thus it can exhibit reasonable results which
are useful for us to better understand the behaviors of many real
networked dynamical systems, such as the evolution of social
networks \cite{Evo:Community}, and the evolution of food webs
\cite{foodweb:weak}.

In this paper, we only investigate the particular case with two
groups, i.e., $h=2$. In fact, the above analysis can be conveniently
generalized to a general case with $h$ groups if we replace $2$ by $h$
in the sine function and the exponential function of  Eq.
(\ref{twoosc}), Eq. (\ref{multipleorder}), Eq.
(\ref{localparameter}) and Eq. (\ref{mmodel}). For the case of a
two-oscillator system, the stable equilibrium states of the phase
difference are $\Delta\theta^{*}=2q\pi/h \pm
\arcsin|\frac{\Delta\omega}{2\gamma w}|/h$ ($q=0,1,2,..., [h/2]$),
where the symbol $[x]$ means taking the integer part of the real
number $x$. For the case of a any-oscillator system, the
equilibrium values of the phase difference are
$\Delta\theta^{*}_{mn}=\frac{2q\pi}{h}\pm
\frac{2|\Delta\omega_{mn}|}{h\gamma(s_m+s_n)}$ ($q=0,1,2,...,
[h/2]$).  Of course, the mechanism of changing the connection strengths
in Eq. (\ref{linkmodel}) should be modified accordingly for the case
of $h>2$, e.g., $\frac{dw_{mk}}{dt}=S(\beta) \epsilon
w_{mk}e^{-h|\Delta\theta_{mk}-\alpha|}|\sin(h\Delta\theta_{mk})|$,
where $\beta=[h\Delta\theta_{mk}/\pi]$,
$\alpha=\{\beta+[1+(-1)^\beta]/2\}\pi/h$, and $S(0)=1$ or
$S(\beta>0)=-1$. Our numerical simulations have verified the
analysis.

In our model, the connection strengths are assumed to respond
immediately to the change of phase difference. Nevertheless,
time-delay inevitably exists in realistic networked systems. For
example, electric signals can only propagate along neural axons at a
finite speed in neural networks. Recently, time delays have been
investigated in some theoretical models of neural networks
\cite{delayedneural} and networked oscillator systems
\cite{REV:Kurmoto}. Interestingly, it is shown that these models can
present very rich dynamical behaviors. We believe that the extension
of our current model to the delay-coupling case may provide more
helpful insights in understanding the coevolution of realistic
networked systems. We keep this problem as our future research
topic.

This work is supported by Temasek Laboratories at National
University of Singapore through the DSTA Project No. POD0613356. SGG
is sponsored by the Science and Technology Commission of Shanghai
Municipality under grant no. 10PJ1403300. SGG is also supported by
the NNSF of China under Grant no. 11075056.


\begin{thebibliography}{10}
\bibitem{explain} This is an approximate value, which dependes the
initial condition of the system.
\end{thebibliography}
\end{document}